\documentclass{aastex}
\usepackage{emulateapj5}
\usepackage{epsfig}
\newcommand{\m}{\mbox}
\newcommand{\Ms}{\m{$\,M_\odot$}}
\newcommand{\Rs}{\m{$\,R_\odot$}}

\newcommand{\kms}{\m{\,km\,s$^{-1}$}}

\newcommand{\xih}{\mbox{[X$_i$/H]}}
\begin{document}

                
         

\title{Formation of the Black Hole in Nova Scorpii}
\author{Ph.\ Podsiadlowski}
\affil{Oxford University, Oxford, OX1 3RH, U.K.}
\authoremail{podsi@astro.ox.ac.uk}

\author{K.\ Nomoto}
\affil{Department of Astronomy \& Research Center for the Early
Universe, School of Science, University of Tokyo, Bunkyo-ku, Tokyo
113-0033, Japan}
\authoremail{nomoto@astron.s.u-tokyo.ac.jp}

\author{K. Maeda}
\affil{Department of Astronomy, School of Science, University of
Tokyo, Bunkyo-ku, Tokyo 113-0033, Japan}
\authoremail{maeda@astron.s.u-tokyo.ac.jp}

\author{T. Nakamura}
\affil{Department of Astronomy, School of Science, University of
Tokyo, Bunkyo-ku, Tokyo 113-0033, Japan}
\authoremail{nakamura@astron.s.u-tokyo.ac.jp}

\author{P.\ Mazzali}
\affil{Osservatorio Astronomico di Trieste, via G. B. Tiepolo 11,
I-34131 Trieste, Italy,
\& Research Center for the Early Universe,
School of Science, University of Tokyo, Tokyo 113-0033, Japan}
\authoremail{mazzali@ts.astro.it}

\author{B.\ Schmidt}
\affil{The Research School of Astronomy and Astrophysics, The Australian
National University, Weston Creek, ACT 2611, Australia}
\authoremail{brian@mso.anu.edu.au}

\begin{abstract}
Israelian et al.\ (1999) showed that the stellar companion of the
black-hole binary Nova Sco is polluted with material ejected in the
supernova that accompanied the formation of the black-hole
primary. Here we systematically investigate the implications of these
observations for the black-hole formation process. Using a variety of
supernova models, including both standard as well as hypernova models
(for different helium-star masses, explosion energies, and explosion
geometries) and a simple model for the evolution of the binary and the
pollution of the secondary, we show that most of the observed
abundance anomalies can be explained for a large range of model
parameters (apart from the abundance of Ti). The best models are
obtained for He star masses of 10 to 16\Ms, where spherical hypernova
models are generally favoured over standard supernova ones. Aspherical
hypernova models also produce acceptable fits, provided there is
extensive lateral mixing. All models require substantial fallback and
that the fallback material either reached the orbit of the secondary
or was mixed efficiently with material that escaped.  The black hole
therefore formed in a two-step process, where the initial mass of the
collapsed remnant was increased substantially by matter that fell back
after the initial collapse. This may help to explain the high observed
space velocity of Nova Sco either because of a neutrino-induced kick
(if a neutron star was formed first) or by asymmetric mass ejection in
an asymmetric supernova explosion.

\end{abstract}


\section{Introduction}
X-ray Nova Sco 1994 (GRO J1655-40; hereafter Nova Sco) is one of the
best-studied black-hole transient of recent years (e.g.,
Bailyn et al.\ 1995; Harmon et al.\ 1995;
Hjellming \& Rupen 1995; Tingay et 
al.\ 1995; Orosz \& Bailyn 1997; van der Hooft et al.\
1998; Shahbaz et al.\ 1999).  It is a low-mass
black-hole binary with an orbital period of $2.61\,$d and relatively
well-determined component masses. The most recent and most
self-consistent analysis of the ellipsoidal light curves of the system
by Beer \& Podsiadlowski (2001) has yielded masses of
$5.4 \pm 0.3\Ms$ and $1.45\pm 0.35\Ms$ for the black hole and the
secondary, respectively, which we adopt in this study\footnote{The studies by 
Orosz \& Bailyn (1997) and van der 
Hooft et al.\ (1998) obtained somewhat higher masses for both components.  
However, both of these studies used a color excess, $E(B-V)\simeq 1.3$, 
that is too large to be consistent with the observed colors of the 
secondary in Nova Sco, assuming that its spectral type is in the range 
of F2-F7 III/IV (see Beer \& Podsiadlowski 2001 for details).}.

Nova Sco stands out among the black-hole transients because of an
unusually high space velocity. Using the $\gamma$-velocity of
Shahbaz et al.\ (1999) and the corrections for Galactic rotation of
Brandt, Podsiadlowski \& Sigurdsson (1995),
Nova Sco's space velocity must exceed $106\kms$, a factor of a few
larger than in any other low-mass black-hole
transient. Brandt et al.\ (1995) concluded that the
most likely explanation for the high space velocity is that the black
hole formed in a two-stage process where the initial collapse led to
the formation of a neutron star accompanied by a substantial kick
(Lyne \& Lorimer 1994). The neutron star was
subsequently converted into a black hole by accretion of matter that
was not ejected in the supernova or a phase transition in the cooling
compact object (Brown \& Bethe 1994).  (For a
different view, see Nelemans, Tauris \& van den
Heuvel [1999].)

That the black hole formed in a supernova event was confirmed
by Israelian et al.\ (1999).  Using
high-resolution Echelle spectroscopy with the Keck Telescope, they
showed that the atmosphere of the secondary was enriched by a factor
of 6$\,$--$\,$10 in several $\alpha$-process elements (O, Mg, Si, S,
Ti; see Table~1). Since some of these elements are almost exclusively 
synthesized during a supernova explosion and cannot have been produced in a
low-mass secondary, the secondary must somehow have been exposed to supernova
material that was ejected when the compact object in Nova Sco was
formed. It should be noted that it is by no means required that the
formation of a black hole is accompanied by a supernova-like event. If
a black hole forms promptly, i.e., on the dynamical timescale of the
collapsing core, very few neutrinos can escape from the collapsing
object (e.g., Gourgoulhon \& Haensel 1993). 
Since delayed neutrino heating may be an essential feature in
producing a successful ejection of the stellar envelope, i.e., a
supernova, a lack of neutrino emission may lead to a ``failed'' supernova in
which the whole star collapses into a black hole. In this case, no
pollution of the secondary is expected, nor is a supernova kick. The
observations of Israelian et al.\ (1999) immediately
rule out a simple prompt black-hole formation scenario for Nova Sco without
an accompanying supernova and suggest that the black hole formed in a 
two-step process, where a neutron star may have formed first and 
was subsequently converted into a black hole by accretion. Alternatively,
a black hole may have formed promptly, but subsequent accretion from
a surrounding disk could have driven a jet-like explosion, as in the
collapsar models of  MacFadyen \& Woosley (1999). In either case, the
compact star could have received a substantial kick, in the first
case due to an asymmetry in the neutrino emission (as may be
the cause of neutron-star kicks), in the latter perhaps because of
an asymmetry in the jets.

The purpose of the present study is to explore the implications of the
observations of Israelian et al.\ (1999) for the
formation of the black hole in Nova Sco in some detail. While Si is
enriched by a factor of 8, Fe -- surprisingly -- is not. Since both
of these elements are produced close to the mass cut above which
matter is ejected in a successful supernova, these observations
directly probe the region that is most crucial in determining whether
a supernova is successful or fails. Indeed, the observations pose an
immediate problem, since, in a standard supernova, the mass cut
($M_{\rm cut}\simeq 1.5\,$--$\,2\Ms$; Thielemann,
Nomoto, \& Hashimoto 1996) is much smaller than the present mass of
the black hole in Nova Sco. Figure~1a shows the composition of the
ejecta for a 16\Ms\ helium star for a standard supernova (i.e., with a
canonical explosion energy $E_{\rm K}=1\times 10^{51}\,$erg; from
Nakamura et al.\ 2001). Elements like S 
and Si, both enhanced significantly in the companion of Nova Sco, 
are produced between the innermost 2.5 and
3.5\Ms\ of the helium core, well below the final mass of the black
hole. There are four possible ways by which some matter synthesized in
this region can reach the secondary and by which the mass of the black
hole can be increased to the present value: (1) by fallback of
material in the supernova explosion (Woosley \&
Weaver 1995), (2) by post-supernova mass transfer from the secondary,
(3) as a result of mixing during the collapse phase and (4) as a
result of a more energetic supernova (a hypernova with a larger mass
cut). The composition of a hypernova model (with $E_{\rm K} = 3 \times
10^{52}\,$erg) is shown in Figure~1b.  It illustrates how, for a more
energetic hypernova, elements such as S and Si are produced much
further out in the core of the helium star. Indeed it was this model
that motivated Israelian et al.\ (1999) to first suggest 
a hypernova model for the formation of the black hole in Nova Sco. This 
and a possible connection to gamma-ray bursts was further developed by
Brown et al.\ (2000).  These various solutions are,
of course, not mutually exclusive, and we shall consider all of them
in the following sections (see Fig.~2 for a schematic picture of the
various cases considered).

\section{Modeling the Pollution in the Secondary}

In order to examine whether the observed pollution of the secondary
is consistent with the predictions of stellar and supernova nucleosynthesis,
we have to follow the evolution of the binary through various evolutionary
phases before and after the supernova and need to model the pollution of
the secondary. In this section we present a very simple model that includes 
the main physical effects. Its purpose is
to serve as a reference model with which we can discuss the physical
implications of the observed abundance anomalies. In \S~3 we critically 
assess some of the assumptions in this model and successively
add physical realism to the model by first including mixing 
in the ejecta (\S~3.3, 3.4) and then considering an aspherical explosion 
(\S~3.5). As the model becomes more realistic, the modeling uncertainties
also increase.

\subsection{Description of the model without mixing (Case A)}

The calculation of the pollution of the secondary requires stellar
models which have been subjected to explosive nuclear burning in a
supernova event. We take these from the library of models calculated
by Nomoto et al.\ (1997, 2001a,b).
 At the time of the explosion, the immediate supernova
progenitor had to be a helium star (or Wolf-Rayet star) in order to
fit into the tight binary orbit implied by the present orbital period
of Nova Sco. We consider helium stars with initial helium-star masses,
$M_{\rm He}^0$, of 6, 8, 10 and 16\Ms. These correspond to
main-sequence masses of $\sim$ 20, 25, 30, 40\Ms, respectively. We
consider two classes of explosion models: supernova models that have a
standard supernova explosion energy of 1 foe ($1\,\mbox{foe}\equiv
10^{51}\,$erg) and hypernova models with explosion energies of 8 and
30 foe (for the 10\Ms\ model) and 30 foe (for the 16\Ms\ model). The
6\Ms\ model is a model calculated for SN~1987A, while the 30 foe,
16\Ms\ model is appropriate for the prototype hypernova SN 1998bw
(Iwamoto et al.\ 1998; 
Nakamura et al.\ 2001)\footnote{The term `hypernova' was coined
by Paczy\'nski (1998) as a model for gamma-ray bursters,
linking them to the cataclysmic deaths of massive stars and the formation
of black holes, as in the failed-supernova model of Woosley (1993) 
(for more details see MacFadyen \& Woosley 1999).
For the purposes of this study, we just define them as very energetic
supernovae with energies  $\ga 10^{52}\,$erg, as has been used by 
Iwamoto et al.\ (1998) and Nomoto et al.\ (2001a,b).}.

Helium/Wolf-Rayet stars are known to lose a substantial fraction
of their envelopes in a stellar wind. To take this into account, we
assume that the helium star has lost an amount $\Delta M_{\rm He}
= g\, (M_{\rm He}^0-M_{\rm BH}^0)$ before the explosion, where 
$M_{\rm BH}^0$ is the initial mass of the compact remnant (neutron
star or black hole). We use the parameter $g$ to vary the total amount 
of wind mass loss before the supernova. 

At the time of the explosion, the masses of the primary and secondary
are $M_{\rm He}$ and $M_2^0$, respectively. When the primary
collapses, it first forms a compact remnant of mass $M_{\rm
BH}^0$. The rest of the envelope is assumed to be ejected initially,
but part of it ($M_{\rm fallback}$) will subsequently fall back,
either because it did not achieve escape velocity or was pushed back
by a reverse shock in the envelope (see Woosley \&
Weaver 1995).  The fallback matter increases the mass of the compact
remnant to $M_{\rm BH}^1$ (indeed, it may be this fallback that leads
to the conversion of the compact remnant into a black hole). Figure~3a
schematically illustrates the definition of these various mass
parameters.

In the simple model we consider first, we assume that {\it all\/} of
the matter that falls back has moved beyond the position of the
secondary (in \S~3.2 we shall critically assess this assumption).
Therefore, the secondary can be polluted {\it twice} with
supernova material, first by all the material that is ejected and then
by material that falls back. We assume that the fraction of matter
that is captured is given by the geometric fraction of the secondary
($[R_2^0/2 a_0]^2\simeq 0.01\,$--$\,$0.03, where $R_2^0 =
0.8\Rs\,(M_2^0/\Ms)^{0.8}$ is the radius of the secondary and $a_0$
the initial orbital separation) times some efficiency factor $f$,
where we assume different efficiency factors for matter that passes
the secondary in the initial ejection ($f_{\rm ejection}$) and for
matter that falls back ($f_{\rm fallback}$). The efficiency factors
can be much smaller than 1, for example, if the supernova leads to
stripping of matter from the secondary (Marietta,
Burrows, \& Fryxell 2000 and \S~3.1), or larger than 1 if gravitational
focusing is important. The latter requires that the relative velocity
of the material is less than the escape velocity of the secondary and can
plausibly only occur for fallback material. We also take into account
the pollution of the secondary that has occurred before the supernova
because of the capture of wind material by the secondary (where we
assume a capture efficiency of 1).

The matter that is captured by the secondary has a much larger mean
molecular weight than the composition of the secondary, a relatively
unevolved star at this stage.
This is secularly unstable and leads to thermohaline mixing in the
secondary (e.g., Kippenhahn, Ruschenplatt \&
Thomas 1980). Since the time scale for thermohaline mixing is short
compared to the evolutionary time scale of the secondary, we assume
that the material captured by the secondary is completely mixed
with the rest of the star after the supernova.

In order to be able to follow the post-supernova evolution, we assume
that the pre-supernova system is circular and that the supernova
explosion is spherically symmetric in the frame of the primary. It is
then straightforward to estimate the post-supernova parameters of the
system (we follow Brandt
\& Podsiadlowski 1995, but for other equivalent treatments, see, e.g.,
Bhattacharya \& van den Heuvel 1991;
Nelemans et al.\ 1999). The eccentricity of the post-supernova binary is
given by
$$ e = \frac{\Delta M_{\rm SN}}{M_{\rm BH}^1+M_2^{0}},$$ where $\Delta
M_{\rm SN}\equiv M_{\rm He}-M_{\rm BH}^1$, the post-supernova semi-major
axis by
$$a_{\rm PSN}=\frac{a_0}{1-e},$$ where $a_0$ is the initial orbital
separation.  The post-supernova system kick velocity can be obtained from
equation~(2.10) in Brandt \& Podsiadlowski (1995) as
$$v_{\rm sys} = v_{\rm orb}^0\,\frac{\Delta M_{\rm SN}}{M_{\rm
BH}^1+M_2^0}\,\,
\frac{M_2^0}{M_{\rm He}+M_2^0},$$
where $v_{\rm orb}^0$ is the pre-supernova orbital velocity of the
system.  Here we have neglected the small change in the mass of the
secondary due to the capture of ejected material from the primary
(typically $\sim 0.2\Ms$), as well as any kick associated with the
interaction of the supernova blast wave with the secondary (see
Marietta et al.\ 2000).

After the supernova, the binary parameters will continue to evolve. The
system will first re-circularize, obtaining a new orbital separation
$$a_{\rm rec} = a_{\rm PSN}\,(1-e^2).$$
Once the secondary starts to fill its Roche lobe, it will start to
lose mass, of which a fraction $\beta$ will be accreted by the primary,
while the rest will be ejected from the system. We assume that the matter
that is lost from the system carries away the same specific angular 
momentum as the primary (see, e.g., Podsiadlowski 
Rappaport, \& Pfahl 2001). This is appropriate if the mass loss
occurs from a region near the primary, as suggested by the relativistic
jets observed from Nova Sco (Hjellming \& Rupen 1995). 

Even though this model is still relatively simple (for example, it
does not take into account a kick due to an asymmetric explosion), it
still contains a large number of essentially unspecified parameters
($M_{\rm He}^0$, $M_{\rm BH}^0$, $M_{\rm fallback}$, $f_{\rm
ejection}$, $f_{\rm fallback}$, $g$, $\beta$). For given values
of $f_{\rm ejection}$ and $f_{\rm fallback}$, we have sampled all the
other parameters in a fairly systematic and comprehensive fashion,
although we generally do not change the present masses of the Nova Sco
components, but keep them fixed at 5.4\Ms\ and 1.45\Ms, respectively\footnote{
We have also performed some calculations using masses of 6\Ms\ and 2\Ms,
respectively, obtaining similar results.}.  In
practice, we proceed in the following way.  For each of the 7
supernova models (i.e., each combination of helium star mass and
explosion energy), we systematically vary the initial black-hole
mass, $M_{\rm BH}^0$, the fallback mass, $M_{\rm fallback}$, 
the wind-loss parameter, $g$, and the mass-accretion parameters, $\beta$ (the
latter two are varied from 0 to 1). Having fixed these parameters, we
can use the present orbital period and masses to reconstruct the
pre-supernova masses and pre-supernova orbital period using the
formalism outlined above. If this reconstruction shows that the radius
of the pre-supernova secondary is smaller than its Roche lobe and that
the system remains bound in the supernova explosion (if $e<1$), we
calculate the pollution and mixing in the secondary for our assumed
values of $f_{\rm ejection}$ and $f_{\rm fallback}$.   In order to
decide whether an acceptable model has been found, we define a quality
parameter as
$$Q = \frac{1}{7}\,\sum_{i=1}^7 \left(\frac{\xih-\xih^{\rm obs}}
{\Delta\,\xih^{\rm obs}}\right)^2,$$ where \xih\ are the
calculated logarithmic abundances (relative to solar) of N, O, Mg, Si,
S, Ti, and Fe and $\xih^{\rm obs}$ are the abundances obtained by
Israelian et al.\ (1999) and $\Delta\,\xih^{\rm
obs}$ are the observational errors (see Table~1). We consider a
particular model acceptable if $Q\le 1$.

\subsection{Results}

In Table~2a we present the results for three combinations of
$f_{\rm ejection}$ and $f_{\rm fallback}$ ([1,1], [0,1], [0,2]), for
all supernova models that produced acceptable fits.
The first two columns specify the
initial mass of the helium star and the supernova explosion energy,
while the next 10 columns give the mean logarithmic abundances
(relative to solar) of He, C, O, Ne, Mg, Si, S, Ca, Ti, and Fe for
acceptable models. The next 6 columns contain the mean values of selected
model parameters (the mass of the He star just before the supernova, 
the initial mass of the compact star, the black-hole
mass after fallback, the initial mass of the secondary, the system
kick velocity and the maximum system kick velocity for each set of
calculations, i.e., each of the 7 supernova models). The last figure
$N_{\rm tot}$ gives the total number of acceptable models for each
supernova set and provides a measure of how easy it is to obtain
an acceptable model for the pollution in the secondary for each
set. All the quoted uncertainties are the standard deviations
calculated for all acceptable models in each set. They are not proper
statistical error estimates, since they are based on an even sampling 
of the unknown parameters.  Nevertheless they give an indication of the range
of the corresponding model parameters.

\noindent{\it Consistency with nucleosynthesis calculations}

As the table shows, many acceptable fits can be obtained for many
plausible combinations of the parameters. In particular, 
acceptable fits can be found for He star models with $M_{\rm He}^0$
of 10 and 16\Ms\ for all explosion energies. To some degree, this
just reflects the fact that explosive nucleosynthesis produces
similar overall abundance patterns in all of these models. Indeed, the overall
consistency of the modelled abundances with the observed pollution in
the secondary of Nova Sco provides confirmation of the general
predictions of stellar and supernova nucleosynthesis. 

The only exception to this picture is the abundance of
Ti, which is too low by at least a factor of 2 in all models. This
may be caused by errors in some of the nuclear cross sections
used in the explosive nucleosynthesis calculation or could provide
evidence for a more a complicated nucleosynthesis environment
in the explosion (see \S~3.5).

In Table~3 we summarize some of the key binary parameters of the best-fit
models for Models A, B and C in Table~2a. Note particularly that 
the initial separation before the supernova is typically less than $6\,\Rs$,
much smaller than the present separation ($15.2\Rs$). This implies that
the secondary was almost filling its Roche lobe before the supernova
and results in a relatively large geometrical cross section for the capture
of material from the supernova ejecta ($\sim 0.26\Ms$). The semi-major
axis of the post-supernova system is only slightly increased (at least
in models where the black hole receives no kick at birth). Hence
there had to be substantial mass transfer ($\sim 1\Ms$) after the supernova 
to widen the system to the present separation.

The model also predicts some other abundance anomalies. The abundance
of C is predicted to be enhanced by up to a factor of 1.7, the abundance
of Ca by a factor of 3 to 7, and the abundance
of Ne by a factor of 2 to 5. The enhancement of Ca is weakly
correlated and the enhancement of Ne is weakly anti-correlated
with the explosion energy (these correlations
are strongest for the 10\Ms\ models). Therefore, these patterns provide,
at least in principle, a means by which one could distinguish between 
a normal supernova and a hypernova event.

\noindent{\it Requirement of fallback}

All models require some fallback, where hypernova models require
the smallest amount. This is the result of two factors: (1) the initial
black-hole mass must be close to the mass cut since matter
near the mass cut (which is much smaller than the present black-hole mass) 
must have been ejected in order to produce the observed abundances.
The mass cut is a function of the mass of the helium star and the energy
of the explosion and is largest for the 16\Ms\ hypernova model
(see Fig.~1).
(2) The black-hole mass cannot increase by much more than $\sim 1\Ms$ 
by mass accretion
from the secondary after the supernova. This is a consequence of the
constraints on the pre-supernova orbital parameters imposed by the
present orbital separation and component masses. A larger amount of
post-supernova mass transfer implies a tighter pre-supernova binary (because
mass transfer widens the system).
The largest amount of mass transfer is therefore determined by the condition
that the pre-supernova secondary just fills its Roche lobe.
We note that this constraint would be substantially weaker if we had allowed 
for an asymmetric supernova explosion.

\noindent{\it Low system kick velocities}

The typical system kick velocities are relatively low, varying from 10
to 60\kms. There are a few extreme cases for the 16\Ms\ models where
the system velocity is as high as $\sim 90\kms$. However, these cases
are very rare and require very special model parameters.  These are
cases where the helium star loses relatively little of its envelope in
a stellar wind before the supernova ($M_{\rm He}\ga 13\Ms$) and where
the system becomes almost unbound in the supernova event (the
immediate post-supernova eccentricity is close to 1). Whether such a
small amount of wind mass loss is physically reasonable is somewhat
doubtful. In any case, since the parameter range that leads to these
high kick velocities is extremely limited, such solutions, while not
formally ruled out, are statistically not favoured. The minimum
observed space velocity of Nova Sco is $106\kms$ and possibly much
larger, since this corresponds to the radial velocity only.  It
therefore appears very unlikely that a symmetric supernova explosion
alone could explain the observed velocity. This implies that an additional
kick, e.g., due to an asymmetry in the explosion, is required. While
our model did not take this possibility into account, our results are
consistent with it. The initial masses of the compact remnant, $M_{\rm
BH}^0$, for many of the helium-star models with a mass of $10\Ms$ and
for normal supernova explosions are consistent with the maximum mass
allowed for neutron stars. Even when $M_{\rm BH}^0$ is larger than the
maximum mass of a neutron star, a neutron star may have formed first,
since, in our simple model, $M_{\rm BH}^0$, strictly speaking, does
not only include the initial mass of the compact remnant, but also any
fallback material that did not reach the orbit of the secondary.
Therefore, a two-stage process where the collapse first leads to the
formation of a neutron star, accompanied by a supernova kick, which is
subsequently converted into a black hole by fallback may well explain the
observed space velocity (see Brandt et al.\ 1995).

We note that these conclusions are not inconsistent with the findings of
Nelemans et al.\ (1999). Their results also imply that one
requires rather {\it special\/} parameters to explain the observed space
velocity of Nova Sco with a {\it symmetric} supernova explosion
alone. In addition, some of the sets of parameters they present for 
illustration can be ruled out by the present investigation which takes 
into account more observational constraints and the reduced masses
of the Nova Sco components.

\section{Discussion}

In the previous section we showed that a relatively simple model
can explain most of the observed  $\alpha$-element enhancements in
the secondary of Nova Sco. We now turn to an examination of the question
whether some of the assumptions that went into the model are
actually physically reasonable.

\subsection{Capture efficiency}

Obviously, a significant fraction of the supernova ejecta must have
been captured by the secondary. This is not necessarily expected.
Marietta et al.\ (2000) recently 
examined the effects of 
a supernova blast wave on a binary companion in the context of a Type Ia
supernova. They showed that a significant part of the outer part
of the secondary was lost by ram-pressure stripping. Instead of accreting 
matter from the supernova ejecta, the secondary
actually lost mass. On the other hand, the layers in which 
$\alpha$-process elements are synthesized are buried deep inside the core and
move with a much lower velocity than the outer layers of the helium
star, which are mainly responsible for the ram-pressure stripping.
This makes it easier for some of that material to 
be captured by the secondary (see the discussion in 
Marietta et al.\ 2000).

The problem is less severe if the material captured comes mainly from
fallback since this material will generally have a much lower
velocity than the escaping material. Indeed, if the velocity relative
to the secondary is smaller than the escape velocity at the surface of
the secondary, gravitational focusing could increase the capture
efficiency above a value of 1.

\subsection{The requirement of fallback}

Substantial fallback is commonly found in supernova explosions with
hydrogen-rich envelopes (see Woosley \& Weaver
1995).  This fallback occurs either because matter (typically just
above the mass cut) did not attain escape speed in the supernova or,
more significantly, because it was pushed back by hydrodynamical effects, 
in particular by a reverse shock that forms at the interface of the He/H
boundary.
However, in the present model there is no hydrogen envelope to produce
the interface for a strong reverse shock.  A strong reverse shock is
also formed at the CO/He interface as seen in Type Ib supernova models
(Hachisu et al.\ 1991, 1994).  
In the absence of a strong reverse shock from the composition interface, the
natural location for fallback, if it occurs, is close to the original
core and not close to the location of the secondary (as required in
our simple model) (unless the system is still surrounded by a common
envelope, which cannot be completely ruled out, but is {\it a priori}
very unlikely).

\subsection{Fallback with mixing?}

A possible solution to this problem is a modification of our model
where fallback occurs from a region near the core, but is accompanied 
by substantial mixing. Some of this mixed material, enriched with 
$\alpha$-element-rich material near the mass cut, is ejected 
and produces the observed pollution in the secondary, while the
rest falls back onto the compact remnant (see Fig.~2, Case B).

There are several scenarios in which such mixing could arise.  (1) In
some of the collapsar model studied by MacFadyen, Woosley \& Heger
(2001), the initial collapse leads to the formation of a neutron star
and a weak supernova shock which fails to eject all of the helium
core. As the shock stalls, matter falls back onto the core converting
the neutron star into a black hole. The energy that is associated with
this collapse may then be able to eject the outer envelope (perhaps in
the form of jet-powered shocks; for a similar suggestion for Nova Sco,
see Brown et al.\ 2000).  Vigorous mixing is expected to occur in this
scenario, since, as material tries to fall back, it is being pushed up
by a low-density neutrino-heated bubble, which is Rayleigh-Taylor
unstable (see, e.g., Kifonidis et al.\ 2000).

(2) If a black hole forms promptly, mixing in the ejecta may be
induced by Rayleigh-Taylor instabilities at the Si/O interface
(Kifonidis et al.\ 2000; Kifonidis 2001) as well as the CO/He
interface (Hachisu et al.\ 1991, 1994).  After mixing, the inner part
of the mixed layers may fall back, increasing the mass of the initial
black hole.  Umeda \& Nomoto (2001) have shown
that such mixing followed by fallback can explain the large Zn
abundance observed in very metal-poor stars and suggested that this may
be a generic feature of core-collapse supernovae, either involving 
the formation of a black hole or a neutron star.

3) In an aspherical explosion, heavy elements synthesized in a
deep layer may be mixed into outer layers in the form of a jet
(Maeda et al.\ 2001), while fallback occurs from the equatorial region.
This is equivalent to the mixing \& fallback process discussed in
Umeda \& Nomoto (2001) and could occur for both the formation of a black
hole or a neutron star.

Such a scenario has several advantages: (1) a neutron star may form
first, and the associated neutrino emission can be responsible for a
standard neutrino-induced supernova kick to explain the observed space
velocity. (2) Material near the mass cut is mixed into the envelope
and can escape with it, explaining the abundance anomalies in the
secondary.  (3) The supernova explosion could be weaker and
ram-pressure stripping would then be less of a problem; this could
allow for more efficient capture of supernova material by the
secondary (even in the case of a hypernova explosion, the ejecta in
the equatorial plane could be moving relatively slowly).

\subsection{Case B: Fallback with mixing simulations}

In order to simulate a fallback model with mixing, we modified the
model described in \S~2.1 by assuming that the layer between $M_{\rm
BH}^0$ and $M_{\rm BH}^0+\Delta M_{\rm mix}$ is completely mixed
during the collapse phase, where we defined $\Delta M_{\rm mix} \equiv
m\,M_{\rm fallback}$ (see Fig.~3b). The definition of $M_{\rm BH}^0$
includes both the initial mass of the compact remnant and any fallback
material that was not mixed with the material that is ejected (this
could, for example, be material from the inner parts of the disk that
forms around the initial compact core, as in the collapsar models of
MacFadyen \& Woosley (1999) and MacFadyen et al.\ (2001) or material
falling back from the equatorial region of the star in an aspherical
explosion (Maeda et al.\ 2001).

In Table~2b we present the results of these
simulations for $m = 1$, 2, 3, and 4. Most of the results are similar
to the earlier calculations: the amount of fallback is similar, 
while the system velocities tend to be a bit larger. One significant
difference is that, with the inclusion of mixing, hypernova models
are preferred over standard supernova models, which tend not
to produce enough S.

The size of the mixing region outside the initial compact remnant
varies quite significantly between different models; in some hypernova
models, the mixing region contains less than 1\Ms, while in normal
supernova models at least several \Ms\ are required.

\subsection{Case C: Non-spherical models}

So far, we have assumed that the ejection of matter occurs in a
more-or-less spherically symmetric way. However, this is not generally
expected for hypernova or collapsar models, in particular those
associated with gamma-ray bursts (MacFadyen \& Woosley
1999). In these models it is generally believed that the core is
initially rapidly rotating and that the accretion of matter onto the
compact object occurs through an accretion disk. Maeda
et al.\ (2001) have recently constructed aspherical hypernova models
for the hypernova SN 1998bw, simulating both the hydrodynamics and the
nucleosynthesis in two dimensions for a helium-star model of
16\Ms\ at the beginning of helium burning. They
showed that the chemical composition of the ejecta is strongly 
dependent on direction. In particular, Fe is mainly ejected in
the polar direction, while O and Mg are preferentially ejected near
the equatorial plane.

To simulate the pollution by an aspherical hypernova, we assume that
the secondary is located in the equatorial plane of the helium star
(and the black-hole accretion disk) and that the secondary captures
material that is within an angle $\theta$ of the equatorial plane,
where $\theta$ is the angular radius subtended by the secondary as seen
from the helium star. The results of these simulations (for their best
model C) are presented in Tables~3 and 4 (models indicated by a $^{*}$). 
Somewhat surprisingly,
none of our simulations produced acceptable fits (the results shown
were obtained by increasing the acceptance parameter $Q$ from 1 to 2).
This is a direct consequence of the large overabundance of O and Mg near the
equatorial plane. All models produce either an unacceptable
overabundance of O and Mg or an unacceptable underabundance of S and
Si, depending on where the cut-off below which matter can be mixed
into the ejecta occurs. 

However, when modeling the shapes of spectral lines in SN 1998bw,
Maeda et al.\ (2001) found that the fits could be
improved if there was some lateral mixing in the ejecta, i.e., between
material ejected in the equatorial plane and material ejected more
along the jet axis (e.g., due to a shear instability). To test this
possibility we also considered a model where we assumed complete
lateral mixing in the ejecta for the same hypernova model as used
above, i.e., where all the material within given velocity bins was
assumed to be mixed completely.  The results of these simulations are
also shown in Tables~3 and 4 and indicated with a prime (').  In
this case, excellent fits are obtained, as one would have expected,
since this model should approximate the spherical hypernova model used
earlier. Interestingly, the Ti abundance is significantly
enhanced ([Ti/H]$\sim 0.5$) and is marginally consistent with the
observed value ([Ti/H]$\sim 0.9\pm 0.4$). This increased Ti abundance
comes at the price of an increased Fe abundance. While both
the Ti and Fe abundances are consistent with the observed values, the
ratio [Ti/Fe] is still significantly below the observed one ($\sim 0.2$
instead of 0.8).  We emphasize that the assumption of complete lateral
mixing is extreme; at present we can not identify any physical process
that would lead to such a result. We note, however,
that H\"oflich, Khokhlov \& Wang (2001) have shown that the jet 
in an aspherical 
supernova model is decelerated at the H/He interface and that material
spreads laterally, although only to a limited extent. Whether this
could provide a viable model, requires computations with much higher 
numerical resolution. Irrespectively, our results suggest that the enhanced 
abundance of Ti (which results from a mixture of the nucleosynthesis
products of complete and incomplete Si burning; see
Umeda \& Nomoto [2001] for a discussion) could
potentially provide a signature for an asymmetric hypernova.

\section{Conclusions}

The main conclusion of this investigation is that, using standard
supernova models and a relatively simple model for the pollution of
the secondary, we can explain the observed $\alpha$-element enhancements
in the secondary of Nova Sco, confirming standard nucleosynthesis
predictions (apart from the abundance of Ti, which is always too low).

Nova Sco presents a clear case for a two-step black-hole formation
process, where a substantial fraction of the black-hole mass is
the result of fallback. In order for the secondary to be polluted with
material near the mass cut, this fallback material must either have
reached the location of the secondary before falling back or, 
more likely, be mixed during the explosion with material that will escape. 
A two-step black-hole formation process may provide a simple explanation for 
the high space velocity of Nova Sco, since the system may have 
received the same type of neutrino-induced kick as normal neutron
stars are believed to. Alternatively, asymmetric mass ejection in
an aspherical supernova/hypernova model may also contribute to the
observed space velocity.

Our analysis shows that helium star models of 10 to 16\Ms\ are most
probable and that both normal supernova as well as more energetic hypernova
models can explain the abundances observed in the secondary. The majority
of acceptable models are, however, hypernova models, in particular
for the more realistic models that include mixing. 
The one aspherical hypernova model we used only provided an acceptable
fit to the observed abundance anomalies when we assumed that there
was extreme lateral mixing between the ejecta in the equatorial plane 
and in the polar direction. Of course, so far we only tried
one helium-star model with a mass of 16\Ms\ for one particular
explosion energy. It is quite possible that, for a lower-mass model, 
the oxygen produced in the equatorial plane would be lower, while 
still preserving most of the hypernova features without requiring
extreme lateral mixing. The modeling of the pollution for the
aspherical explosion is clearly even more uncertain than in the
spherical case, which suggests that it is important to study mixing
processes in aspherical models with much higher numerical resolution.

On the other hand, it is also worth noting that it is not necessarily
expected that the helium-star progenitor of the black hole was rapidly
rotating, since the progenitor is likely to have passed through an
extended red-supergiant phase before the system experienced a
common-envelope and spiral-in phase. It is quite likely that, in the
red-supergiant phase, the helium core would have been significantly spun down
by both hydrodynamical (Heger, Langer, \& Woosley
2000) and magnetohydrodynamical processes (Spruit \&
Phinney 1998), which efficiently couple the core to the slowly
rotating, convective envelope (as may also be required to explain the
low initial spin periods of the majority of radio pulsars in supernova
remnants). Strohmayer (2001) has provided some
evidence that the black hole may be rotating rapidly at the present
time (based on observations of quasi-periodic oscillations at
450\,Hz). However, as our modeling has shown, the black hole is likely to
have accreted $\sim 1\Ms$ after the supernova from the companion star
(also see Beer \& Podsiadlowski 2001). Since this accretion
would also have spun up the black hole, the observations by
Strohmayer (2001) do not provide a strong
constraint on the rotation of the black hole at birth.

It is interesting to speculate on the relation between Nova Sco and
other low-mass black-hole binaries. Most low-mass black-hole binaries
have rather low space velocities, entirely consistent with the
expected velocity dispersion caused by  scattering by molecular
clouds and spiral arms (Brandt et al.\ 1995).  This
may suggest that the black holes in these systems formed
promptly, i.e., without a kick or significant mass ejection
(see, however, the discussion in Nelemans et al.\
1999 for a rather different point of view). This would then imply
that the secondaries in these systems should not show the same enhancement
in $\alpha$-process elements as the secondary in Nova Sco, a prediction that
should be checked with future high-resolution spectral observations of
these systems. Orosz et al.\ (2001) recently reported
the discovery of similar over-abundances of $\alpha$-process elements in the 
B-star secondary of another X-ray transient J1819.3-2525 (V4641 Sgr).
Interestingly, this systems also appears to have an anomalously
high space velocity (its measured $\gamma$-velocity is $107\kms$). 
Unfortunately, this velocity estimate is quite uncertain at present
because of the proximity of the system to the Galactic center, which
makes the distance estimate very sensitive to the precise distance.
Nevertheless, if confirmed, it would suggest that the system may have 
experienced a similar evolutionary history as Nova Sco.

\eject
\begin{figure*}
\begin{minipage}{\linewidth}
\centerline{\epsfig{file=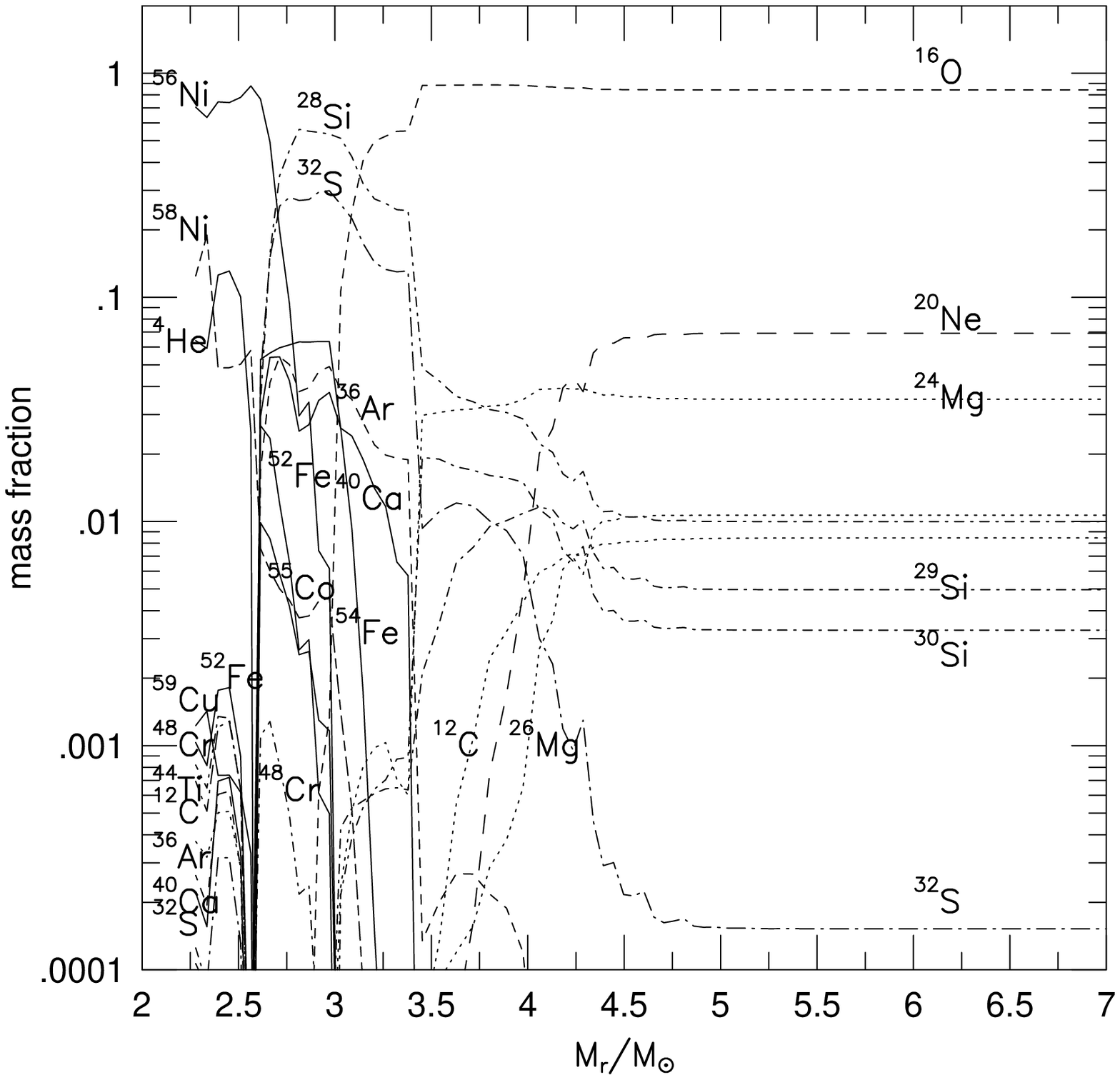,height=0.7\linewidth,angle=0}}
\end{minipage} 
\hfill \\
\begin{minipage}{\linewidth}
\centerline{\epsfig{file=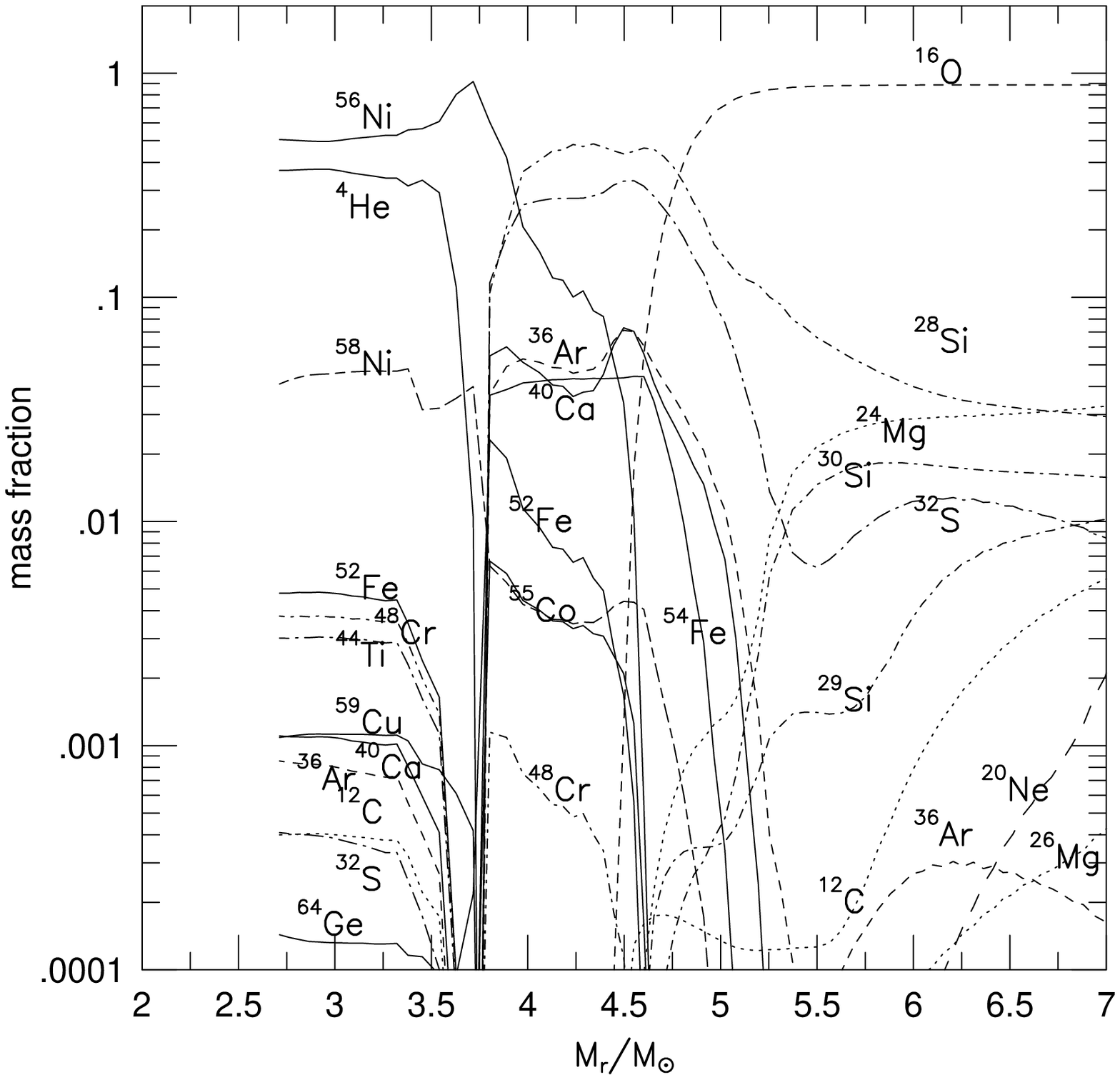,height=0.7\linewidth,angle=0}}
\end{minipage}
\caption{Isotopic chemical composition
for a 16\Ms\ He star for a normal supernova (a) with explosion
energy $E_{\rm K}=1\times 10^{51}\,$erg and a hypernova (b)
with explosion energy $E_{\rm K} = 3\times 10^{52}\,$erg (from
Nakamura et al.\ 1998).}
\end{figure*}
\par
\vfill\eject
\begin{figure*}
\centering
\epsfig{file=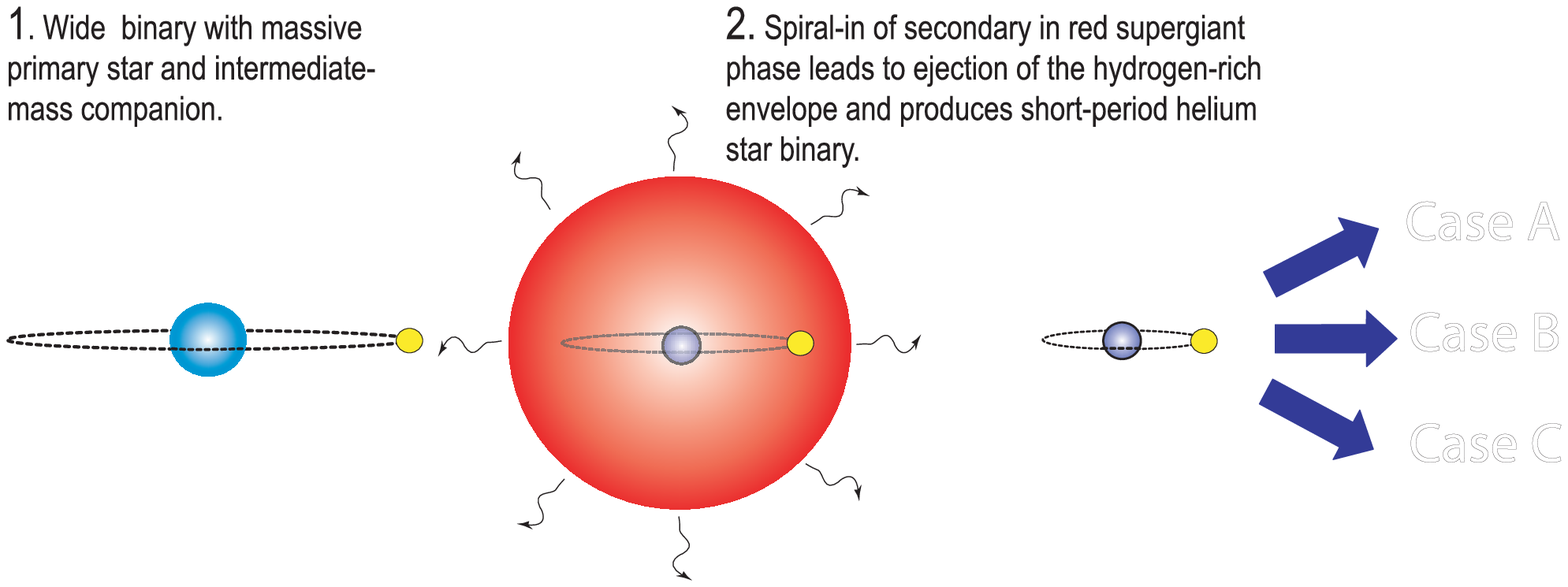,width=6.in,angle=0}
\vspace{1cm}
\epsfig{file=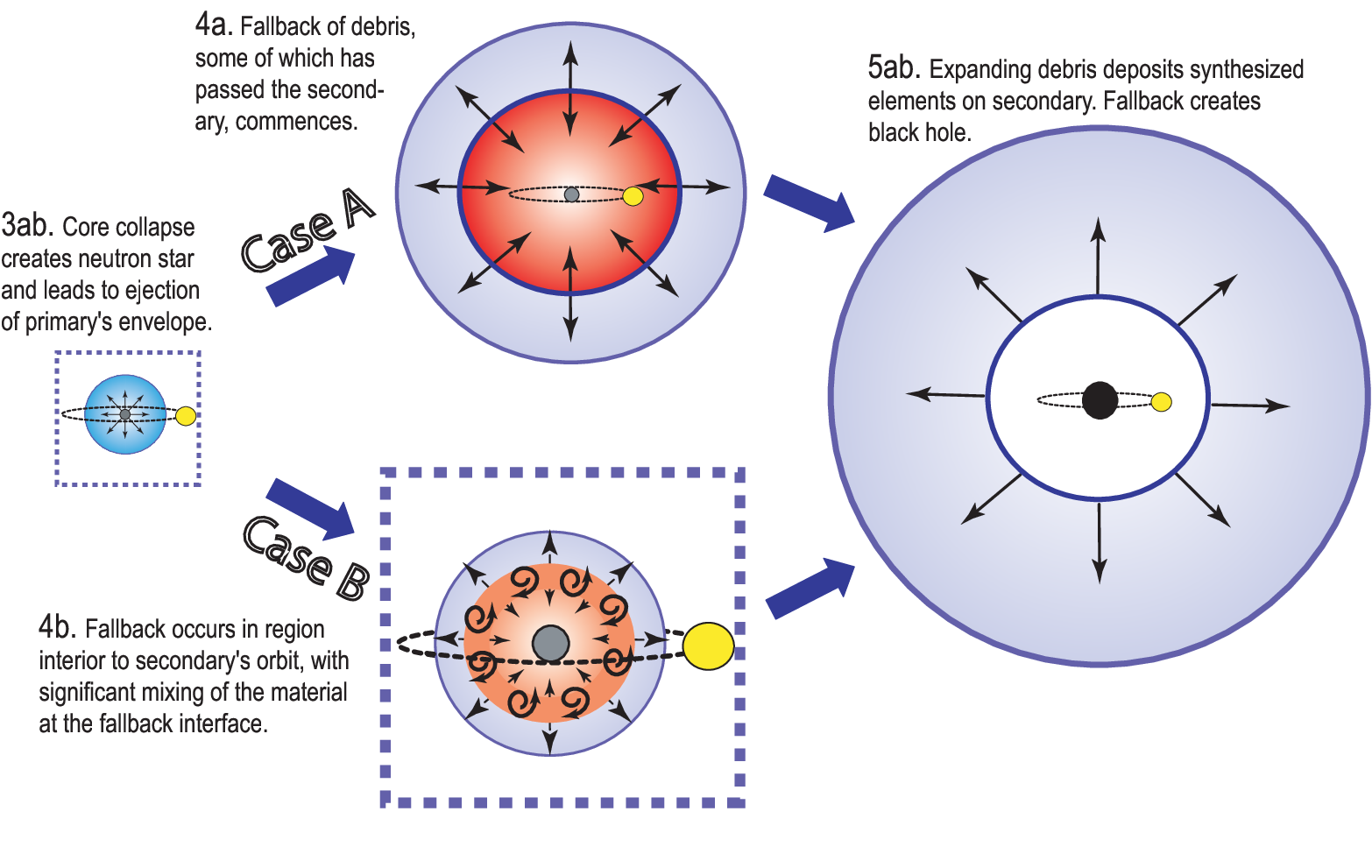,width=6.in,angle=0}
\end{figure*}
\begin{figure*}
\epsfig{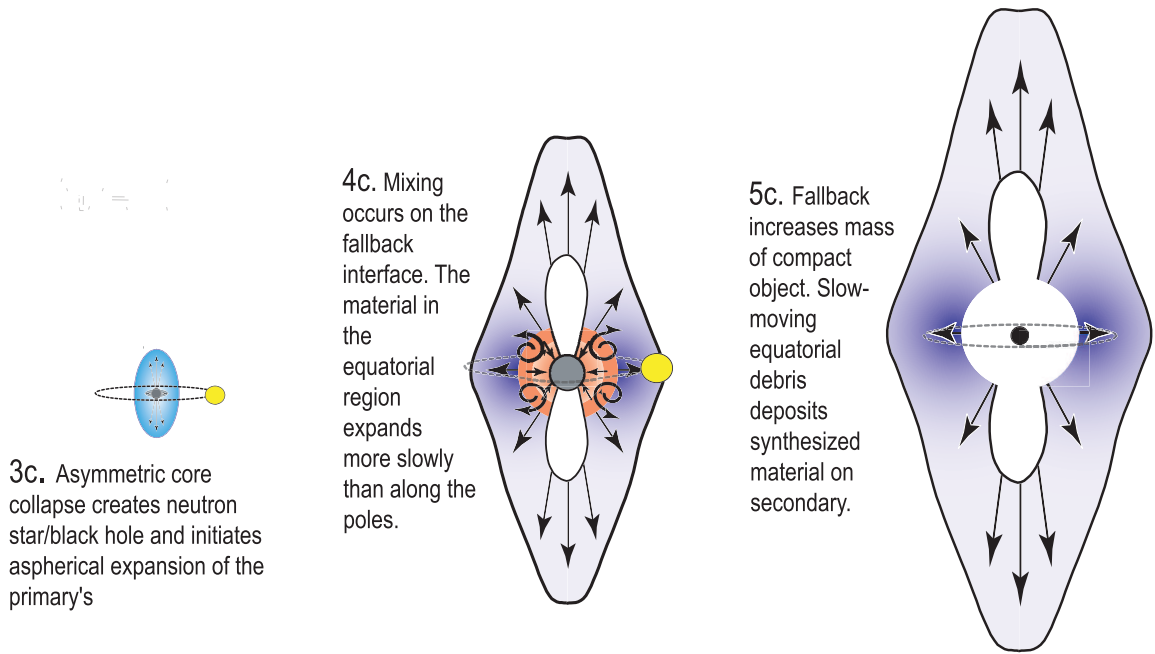}
\caption{Schematic diagram for the various pollution
scenarios considered.}
\end{figure*}
\begin{figure*}
\begin{minipage}[t]{\linewidth}
\centerline{\epsfig{file=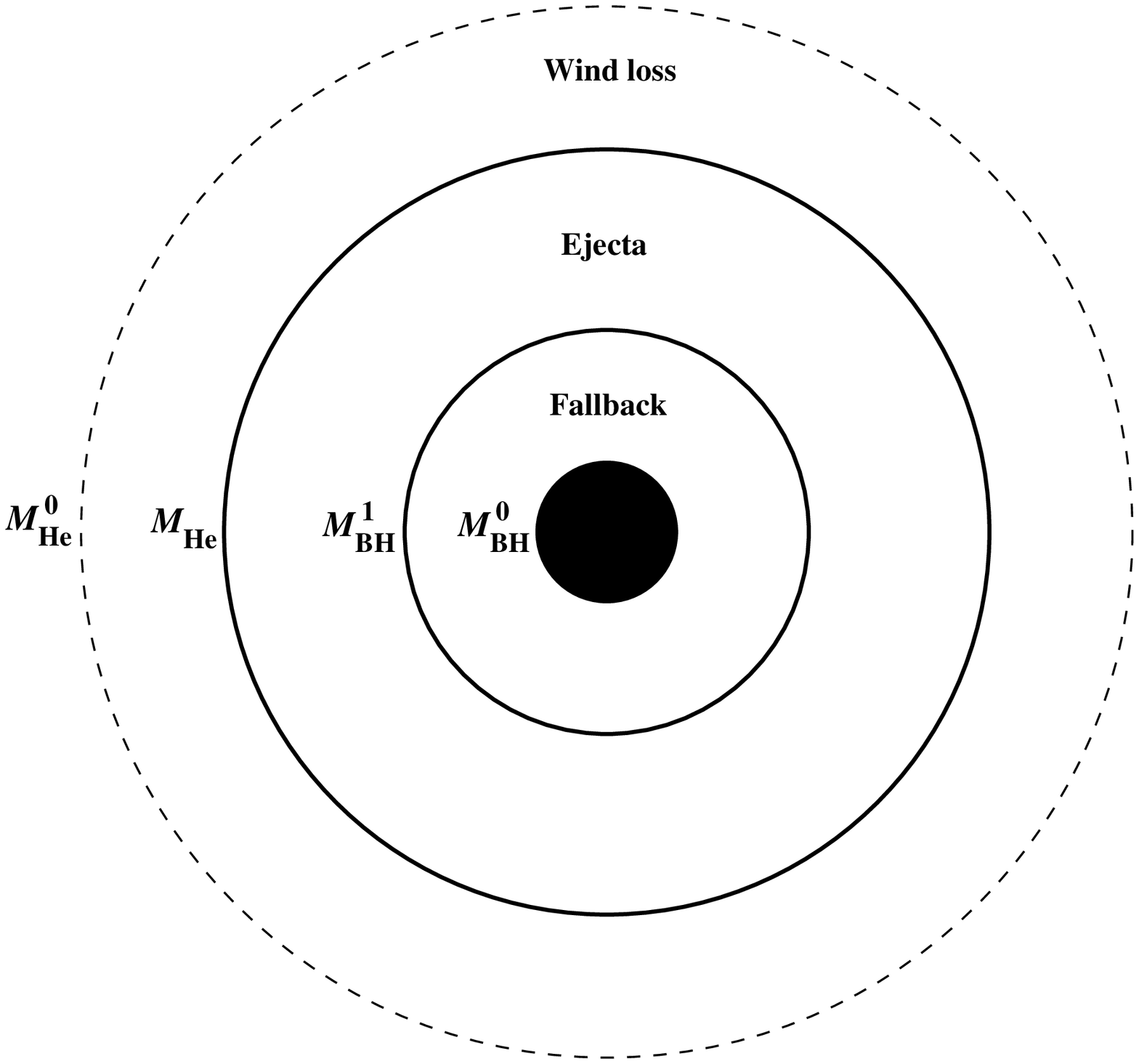,height=0.6\linewidth,angle=-90}}
\end{minipage} 
\hfill \\ \\
\begin{minipage}[t]{\linewidth}
\centerline{\epsfig{file=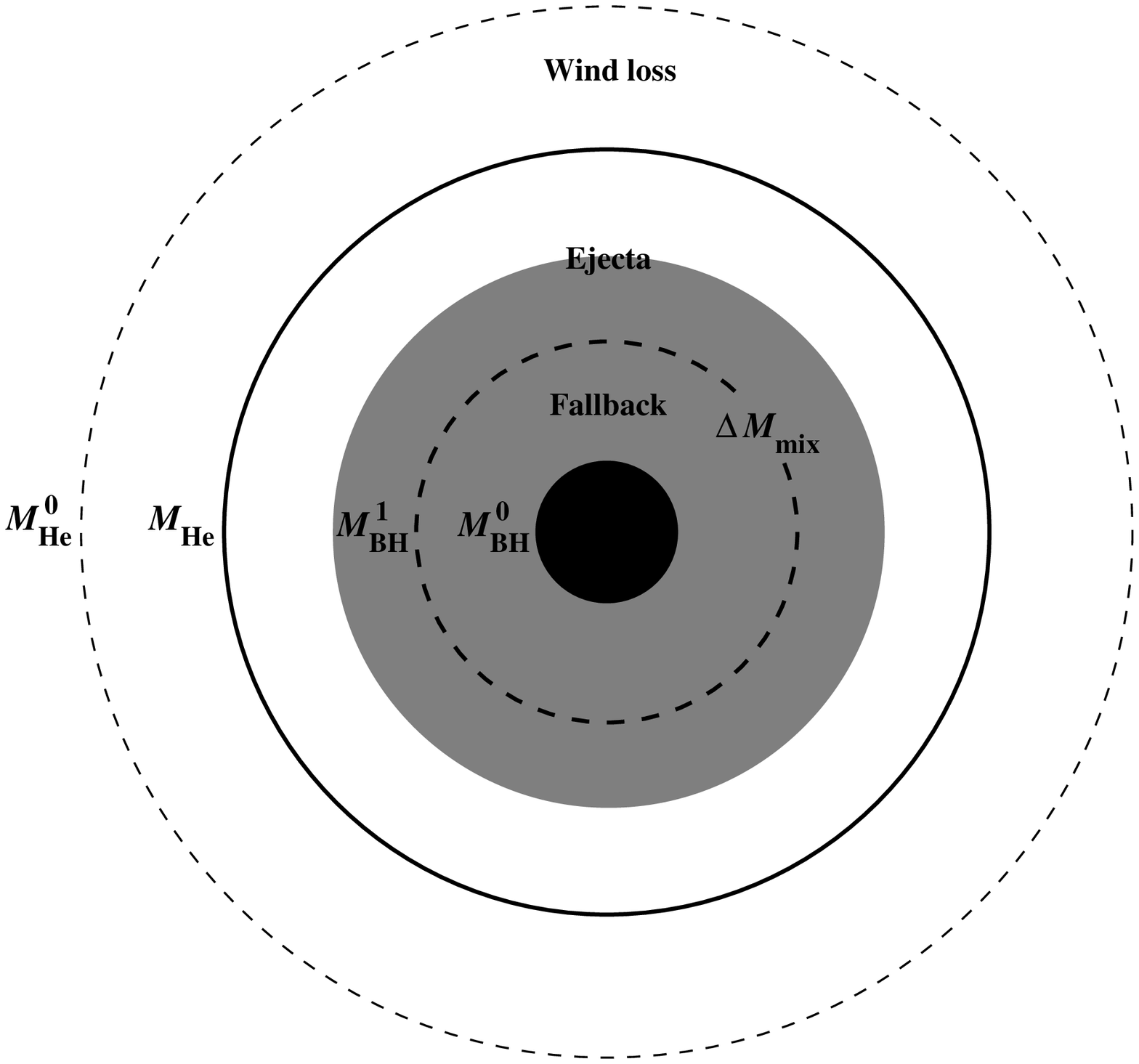,height=0.6\linewidth,angle=-90}}
\end{minipage}
\caption{Schematic diagram defining the various mass 
parameters in the simple fallback model (a) and the fallback with mixing 
model (b). $M_{\rm BH}^0$ is the initial black-hole mass, $M_{\rm BH}^1$
the black-hole mass after fallback, $M_{\rm He}$ the mass of the helium
star at the time of the supernova, $M_{\rm He}^0$ the initial mass
of the helium core (without wind mass loss), $\Delta M_{\rm mix}$ the
mass in the mixing region (b).}
\end{figure*}
\vfill\eject
\begin{figure*}
\centerline{\epsfig{file=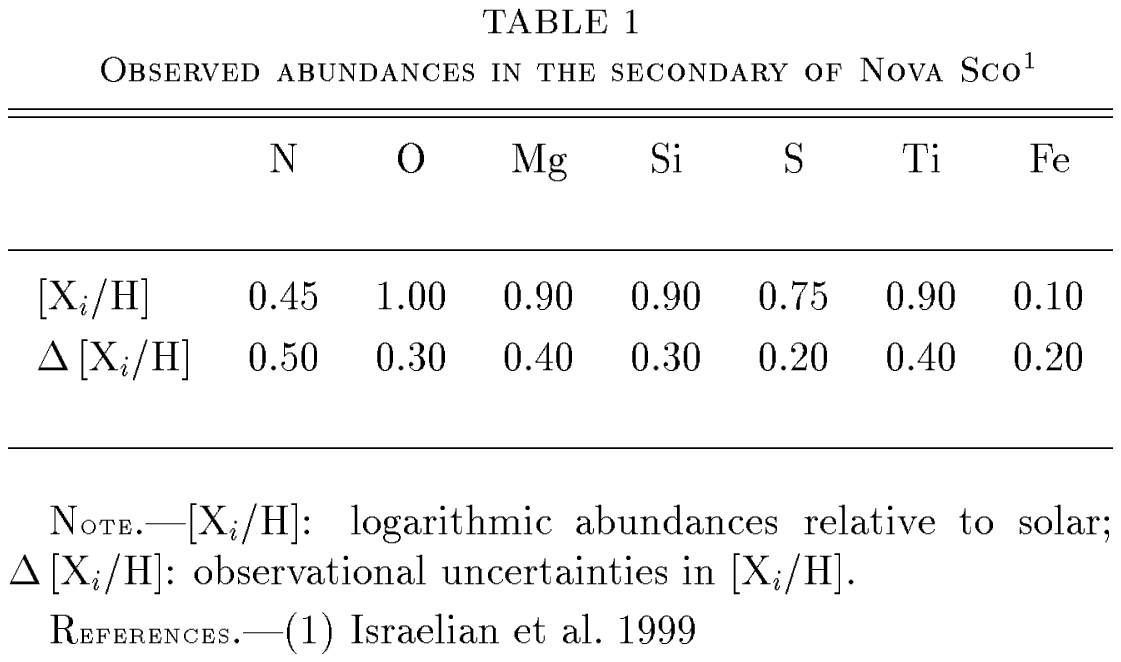,width=\linewidth}}
\end{figure*}
\begin{figure*}
\centerline{\epsfig{file=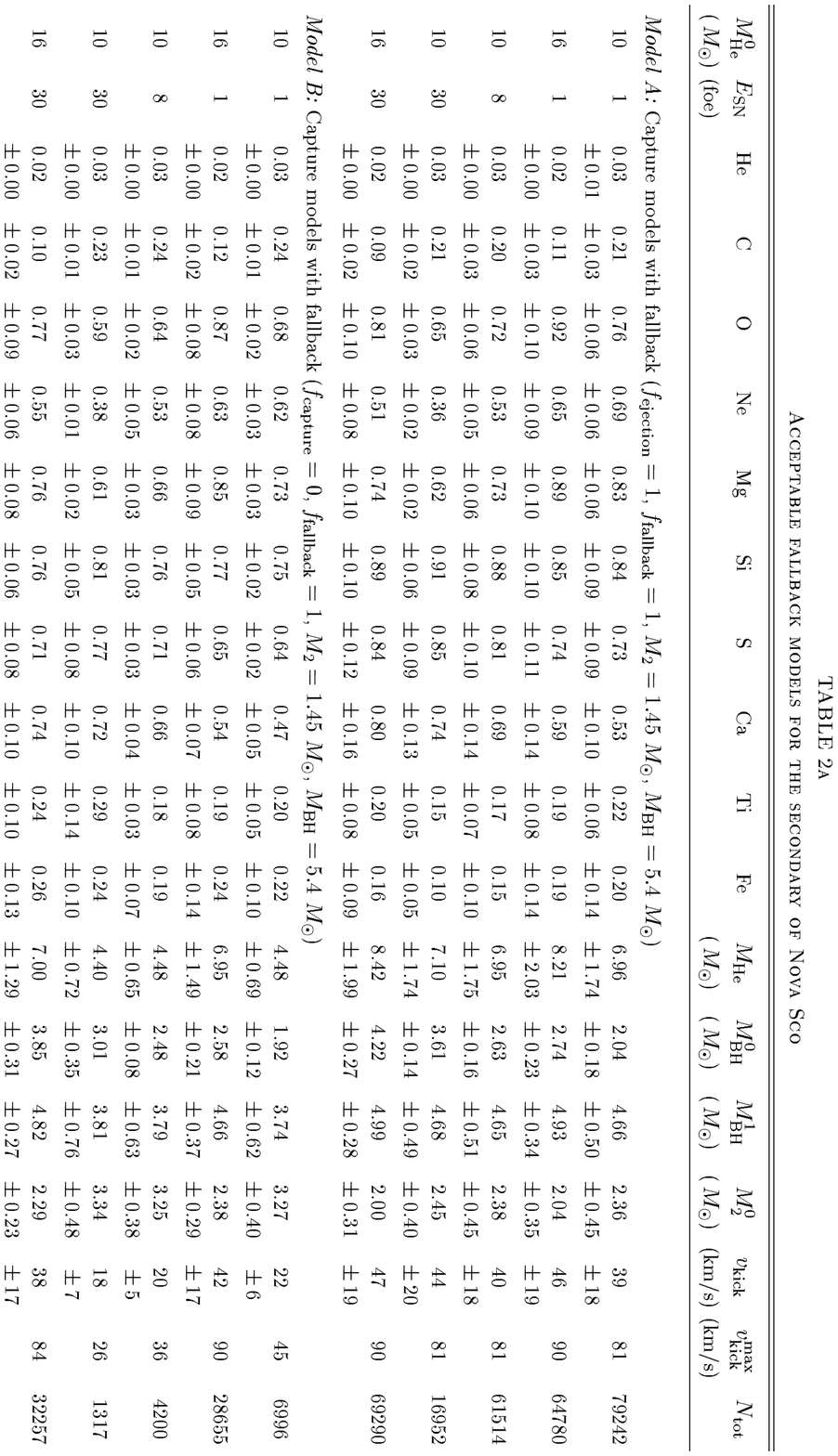,width=\linewidth}}
\end{figure*}
\begin{figure*}
\centerline{\epsfig{file=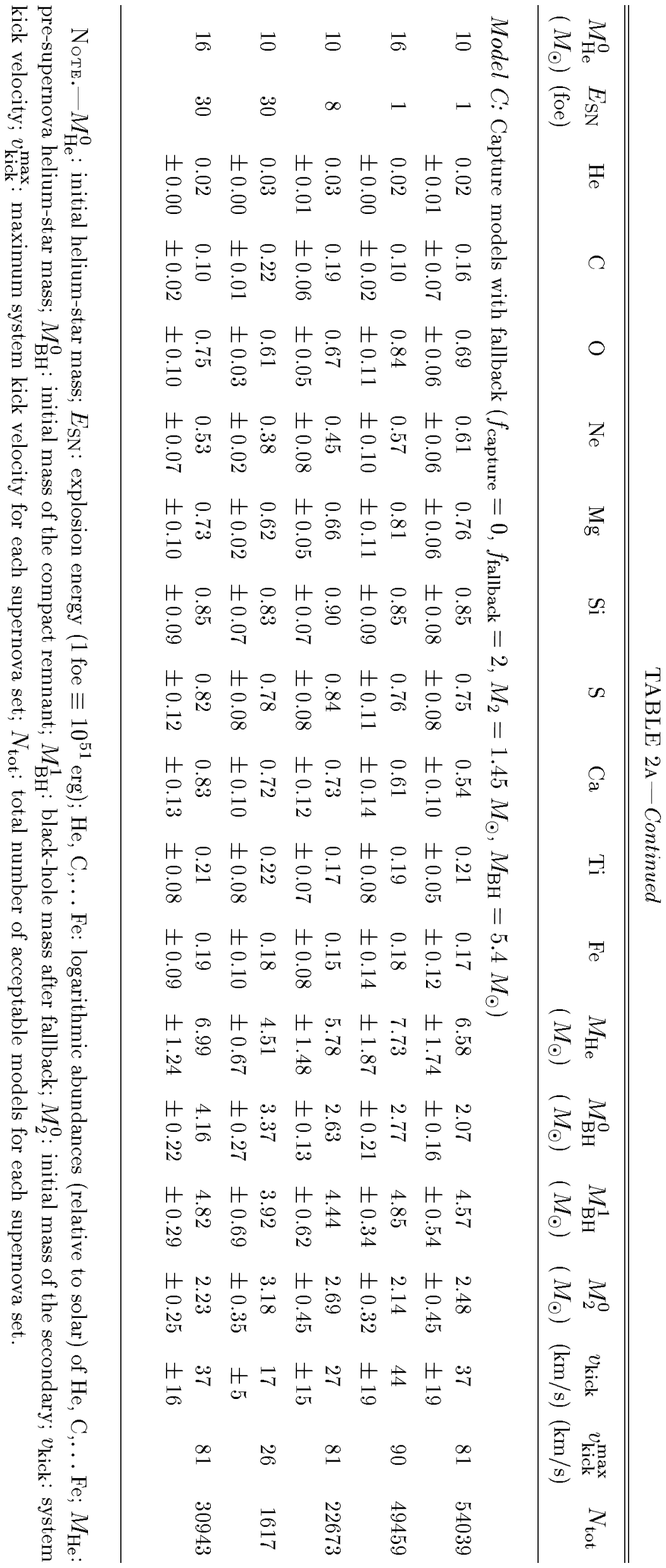,width=\linewidth}}
\end{figure*}
\begin{figure*}
\centerline{\epsfig{file=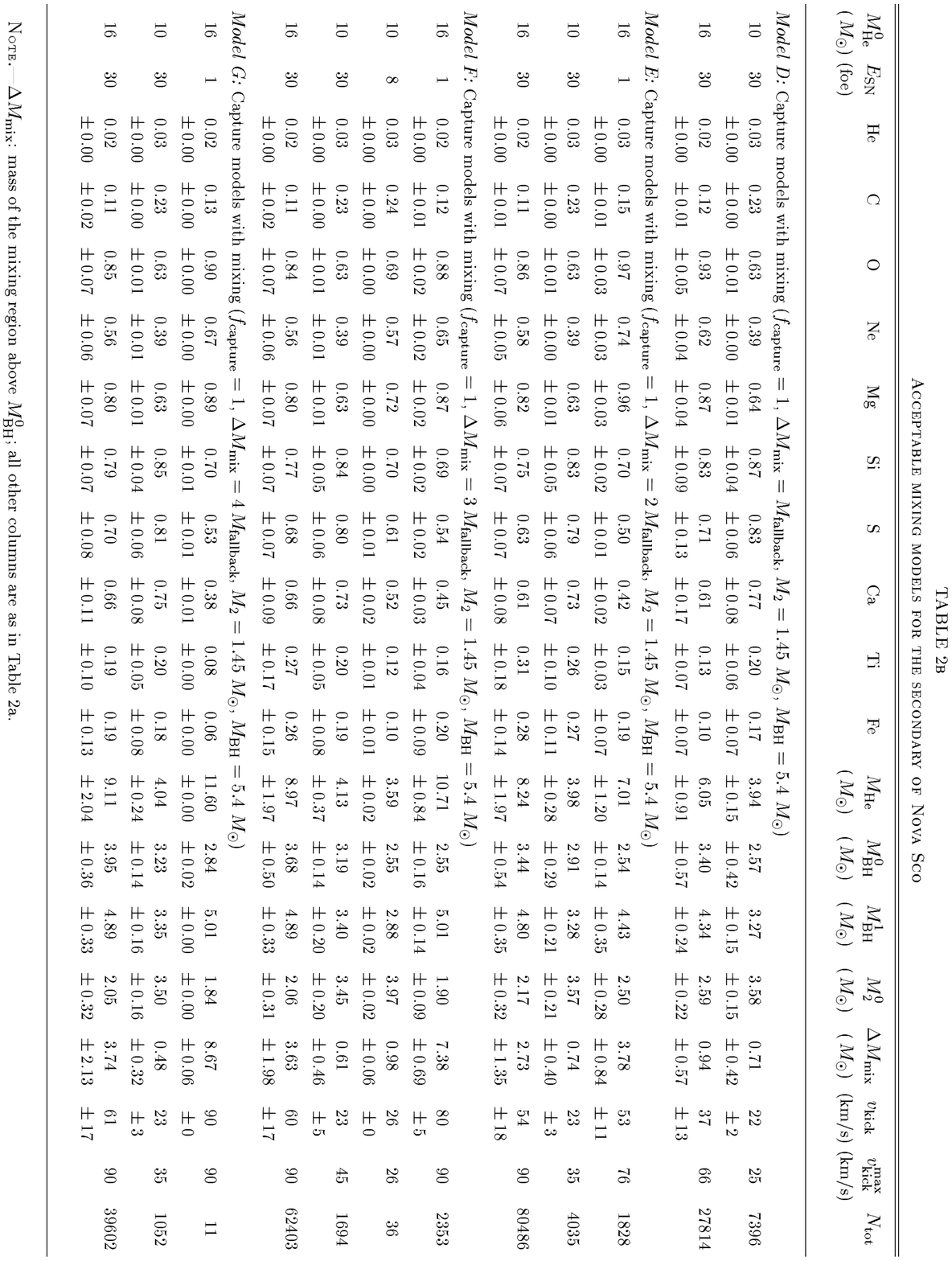,width=\linewidth}}
\end{figure*}
\begin{figure*}
\centerline{\epsfig{file=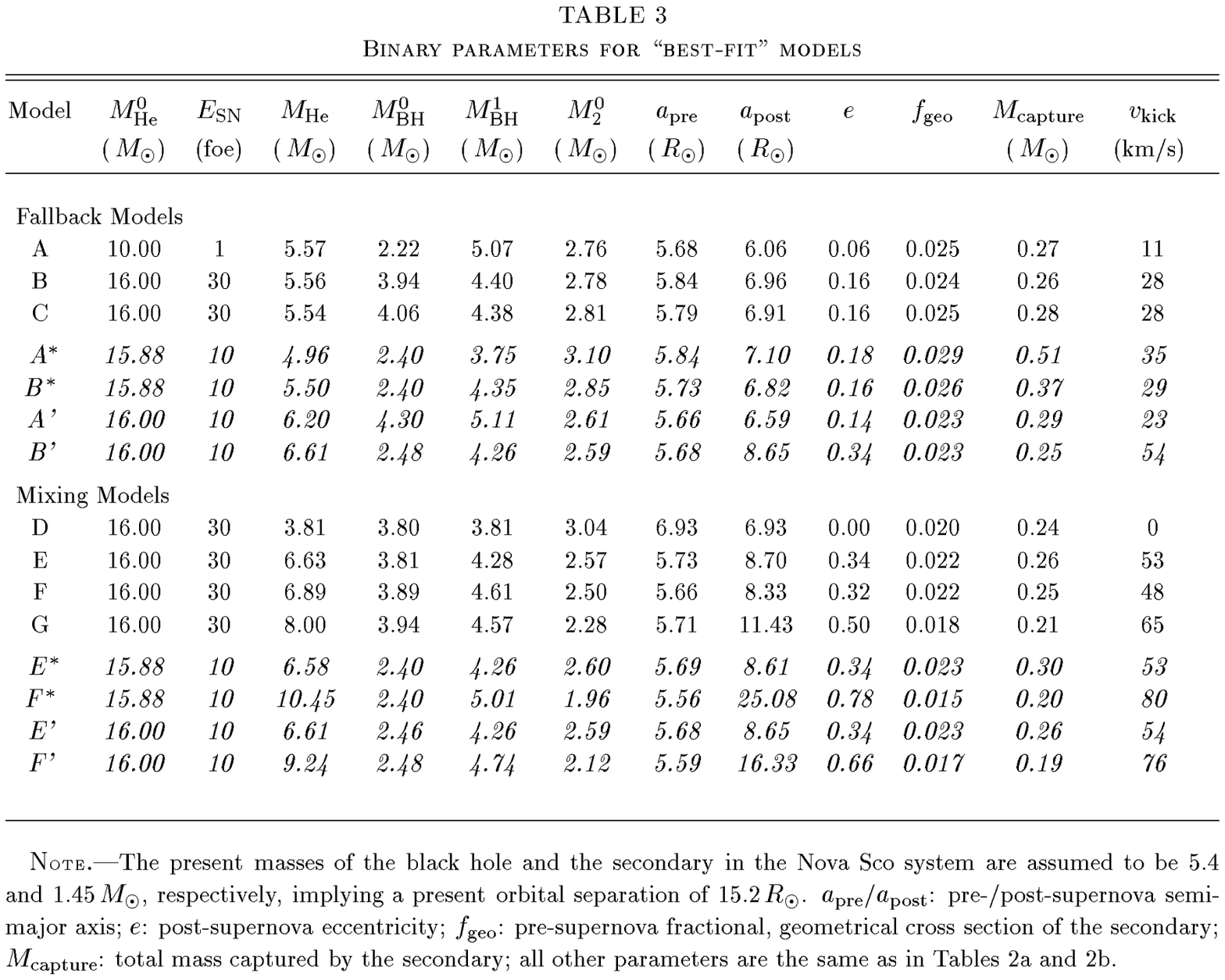,width=\linewidth}}
\end{figure*}
\begin{figure*}
\centerline{\epsfig{file=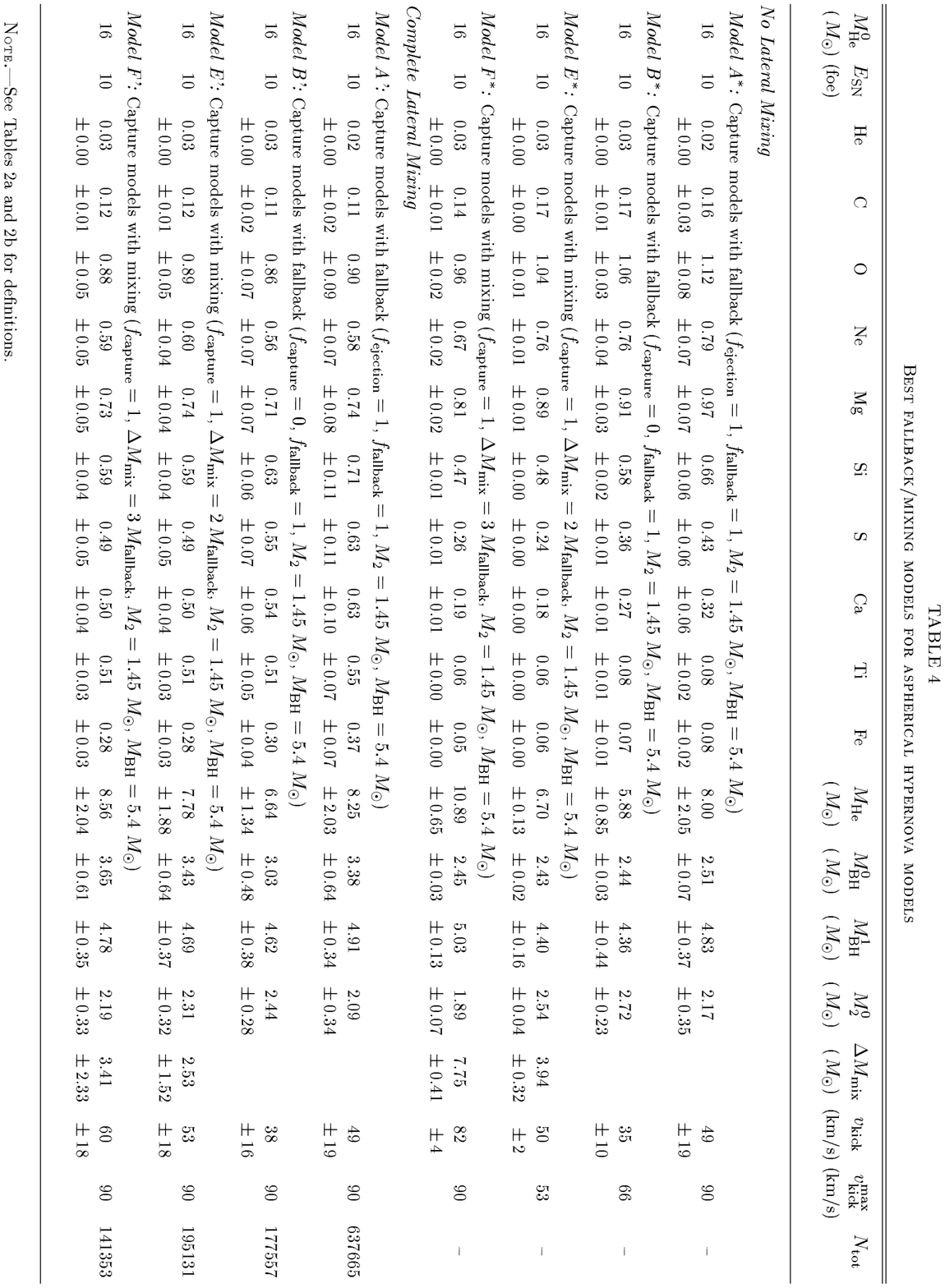,width=\linewidth}}
\end{figure*}
\end{document}